\documentclass[letterpaper,english,prd,preprint,nofootinbib,showpacs,preprintnumbers,fleqn,floatfix]{revtex4}
\usepackage[T1]{fontenc}
\usepackage[latin1]{inputenc}
\usepackage{graphicx}
\usepackage{amssymb}

\makeatletter

\usepackage{babel}
\makeatother
\begin{document}

\newcommand{\prescr}[1]{{}^{#1}}

\newcommand{\ra}{\rightarrow}

\newcommand{\alphas}{\alpha_{s}}

\newcommand\gsim{\mathrel{\rlap{\raise.4ex\hbox{$>$}} {\lower.6ex\hbox{$\sim$}}}}  \newcommand\lsim{\mathrel{\rlap{\raise.4ex\hbox{$<$}} {\lower.6ex\hbox{$\sim$}}}}  

\newcommand\etmiss{E_{T}\hspace{-13pt}/\hspace{8pt}}

\newcommand{\mHQ}{m_b}

\newcommand{\qHQ}{b}

\def\eslt{\not\!\!{E_T}} \def\mslash{\not\!\!{m}} \def\to{\rightarrow} \def\Phat{\hat{\Phi}} \def\pT{$p_{T}$} \def\bi{\begin{itemize}} \def\ei{\end{itemize}} \def\be{\begin{equation}} \def\ee{\end{equation}} \def\bea{\begin{eqnarray}} \def\eea{\end{eqnarray}} \def\te{\tilde e} \def\tl{\tilde l} \def\tu{\tilde u} \def\ts{\tilde s} \def\tb{\tilde b} \def\tf{\tilde f} \def\td{\tilde d} \def\tQ{\tilde Q} \def\tL{\tilde L} \def\tH{\tilde H} \def\tst{\tilde t} \def\ttau{\tilde \tau} \def\tmu{\tilde \mu} \def\tg{\tilde g} \def\tnu{\tilde\nu} \def\tell{\tilde\ell} \def\tq{\tilde q} \def\tw{\widetilde W} \def\tz{\widetilde Z} \def\ttb{t \overline{t}} \def\qqb{q \overline{q}} \def\alt{\stackrel{<}{\sim}} \def\agt{\stackrel{>}{\sim}} \def\X{\times} \def\emis{\not\hskip-5truedd E_{T} } \def\LQ{{L\!\!Q}} \def\tanb{\tan\beta} \def\hbb{{\cal H}(b\bar{b})} \def\mH{\mathcal{H}}

\def\toploop{{\ \rightarrow\hspace*{-0.4cm}^\rhd}\ } \def\bbH {$b\bar{b}\rightarrow H$ } \def\bbA {$b\bar{b}\rightarrow A$ } \def\bbmH{$b\bar{b}\rightarrow \mH$ } \def\ggH {$gg \toploop H$ } \def\ggA {$gg\toploop A$ } \def\ggmH{$gg\toploop \mH$ }

\pacs{12.15.Ji, 12.38 Cy, 13.85.Qk}

\preprint{hep-ph/0603049}

\preprint{ANL-HEP-CP-06-27 }

\title{Heavy-flavor effects in supersymmetric Higgs boson production at
hadron colliders }

\author{Alexander Belyaev$^{1}$, Stefan Berge,$^{2}$ Pavel M. Nadolsky,$^{3}$\\
Fredrick I. Olness,$^{2}$ and C.-P. Yuan$^{1}$}

\affiliation{$^{1}$Department of Physics and Astronomy, Michigan State University,
\\
East Lansing, MI 48824, USA\\
$^{2}$Department of Physics, Southern Methodist University, \\
Dallas, Texas 75275-0175, U.S.A.\\
$^{3}$High Energy Physics Division, Argonne National Laboratory,\\
Argonne, IL 60439-4815, U.S.A. }

\begin{abstract}
We evaluate the effect of the bottom-quark mass on resummed transverse
momentum ($q_{T}$) distributions of supersymmetric Higgs bosons at
the Tevatron and LHC. The mass of the bottom quark acts as a non-negligible
momentum scale at small $q_{T}$ and affects resummation of soft and
collinear radiation in this region. The improved treatment of the
$b$-quark mass and kinematical effects leads to observable modifications
in the resummed predictions for both colliders. 
\end{abstract}

\date{\today{}}

\maketitle

\section{Introduction\label{sec:Introduction}}

Understanding the nature of electroweak symmetry breaking is the central
challenge for high-energy physics. The search for Higgs bosons, assumed
to be responsible for the generation of gauge-boson and fermion masses,
is the primary task for the existing and future colliders.

The Higgs sector may be represented by one complex scalar doublet,
as it is economically realized in the Standard Model (SM), or by two
or more doublets, as it takes place in the Minimal Supersymmetric
Standard Model (MSSM) and its extensions. An important feature of
MSSM is that, for large values of $\tan\beta$, the Yukawa couplings
of the $b$-quarks to the neutral Higgs bosons $\mH$ (where $\mH=h$,
$H$, or $A$) are strongly enhanced compared to the SM $b\bar{b}H_{SM}$
Yukawa coupling. Consequently, production of supersymmetric Higgs
bosons in $b\bar{b}$ fusion can have a large cross section in supersymmetric
extensions of the Standard Model \cite{Carena:2002es,Spira:1997dg,Assamagan:2004mu,Balazs:1998nt,Diaz-Cruz:1998qc}.

The partonic processes contributing to the inclusive Higgs boson production
with enhanced $b\bar{b}\mH$ coupling are represented by (a) $b\bar{b}\rightarrow\mH$;
(b) $gb\rightarrow\mH b$; and (c) $gg\rightarrow b\bar{b}\mH$ scattering.
The three processes (a,b,c) all give rise to the same hadronic final
states, with two $B$-mesons appearing in different, but overlapping,
regions of phase space. The distinction between the three processes
depends very much on the factorization scheme adopted for the QCD
calculation, as has been recently reviewed in Ref.~\cite{Belyaev:2005nu}.
The $\mH$${b\bar{b}}$ processes have been extensively studied recently
in SM and MSSM scenarios~\cite{Dicus:1998hs,Campbell:2002zm,Maltoni:2003pn,Boos:2003yi,Hou:2003fm,Dittmaier:2003ej,Harlander:2003ai,Dawson:2003kb,Kramer:2004ie,Field:2004nc,Dawson:2004sh,Dawson:2005vi}.

As shown in Refs.~\cite{Belyaev:2002zz,Belyaev:2005ct}, the correct
model for the transverse momentum distribution of the Higgs boson
is crucial for unambiguous reconstruction of the Higgs boson mass
in the $\mH\rightarrow\tau\tau$ decay channel. It is also important
for discriminating the signal events from the backgrounds by examining
the $q_{T}$ distribution of the Higgs boson in ${\mH}b\bar{b}$ associated
production, followed by ${\mH}\rightarrow b\bar{b}$ decay~\cite{Carena:2000yx}.
The transverse momentum ($q_{T}$) distributions of Higgs bosons may
be sensitive to the mass $m_{b}$ of the bottom quark when $q_{T}$
is comparable to $m_{b}$. In Refs.~\cite{Berge:2005rv,Belyaev:2005bs}
, we study the effect of the initial-state multiple parton radiation
and heavy-quark masses on the transverse momentum distribution in
the $b\bar{b}\rightarrow\mH$ process. Here we summarize the results
of those two papers.

\section{\label{sec:Formalism}Transverse Momentum Resummation for Massive
Quarks}

The resummed differential cross section for inclusive production of
Higgs bosons in scattering of initial-state hadrons $A$ and $B$
takes the form~\cite{Collins:1985kg}\begin{equation}
\frac{d\sigma}{dQ^{2}dydq_{T}^{2}}=\int_{0}^{\infty}\frac{{\rm b}d{\rm b}}{2\pi}\, J_{0}(q_{T}{\rm b})\, W({\rm b},Q,x_{A},x_{B},m_{b})\,\,+\,\, Y(q_{T},Q,y,m_{b}),\label{WYDY}\end{equation}
 where $y$ is the rapidity of the Higgs boson, $x_{{A,B}}\equiv Qe^{\pm y}/\sqrt{S}$
are the Born-level partonic momentum fractions, $S$ is the square
of the center-of-mass energy of the collider, and $J_{0}(q_{T}{\rm b})$
is the Bessel function. The resummed form factor $W$ is given in
impact parameter ($\mathrm{b}$) space and factorizes as\begin{equation}
W({\rm b},Q,x_{{A}},x_{{B}},m_{b})=\frac{\pi}{S}\,\sum_{j,k}\sigma_{jk}^{(0)}\, e^{-\mathcal{S}({\rm b},Q,m_{b})}\,\,{\mathcal{\overline{P}}}_{j/A}(x_{A},{\rm b},m_{b})\,\,{\mathcal{\overline{P}}}_{k/B}(x_{{B}},{\rm b},m_{b}),\label{WCSS}\end{equation}
 where the summation is performed over the relevant parton flavors
$j$ and $k$. Here, $\sigma_{jk}^{(0)}$ is a product of the Born-level
prefactors, $e^{-{\mathcal{S}}({\rm b},Q,m_{b})}$ is an exponential
of the Sudakov integral \begin{eqnarray}
{\mathcal{S}}({\rm b},Q,m_{b})\equiv\int_{b_{0}^{2}/{\rm b}^{2}}^{Q^{2}}\frac{d\bar{\mu}^{2}}{\bar{\mu}^{2}}\biggl[{\mathcal{A}}(\alpha_{s}(\bar{\mu}),m_{b})\,\mathrm{ln}\biggl(\frac{Q^{2}}{\bar{\mu}^{2}}\biggr)+{\mathcal{B}}(\alpha_{s}(\bar{\mu}),m_{b})\biggr],\label{Sudakov}\end{eqnarray}
 with $b_{0}\equiv2e^{-\gamma_{E}}\approx1.123$, and ${\mathcal{\overline{P}}}_{j/A}(x,{\rm b},m_{b})$
are the ${\rm b}$-dependent parton distributions for finding a parton
of type $j$ in the hadron $A$. In the perturbative region (${\rm b}^{2}\ll\nolinebreak\Lambda_{QCD}^{-2}$),
the distributions ${\mathcal{\overline{P}}}_{j/A}(x,\mathrm{b},m_{b})$
factorize as\begin{eqnarray}
\left.\overline{{\mathcal{P}}}_{j/A}(x,\mathrm{b},m_{b})\right|_{\mathrm{b}^{2}\ll\Lambda_{QCD}^{-2}} & = & \sum_{a=g,u,d,...}\,\int_{x}^{1}\,\frac{d\xi}{\xi}\,{\mathcal{C}}_{j/a}(x/\xi,\mathrm{b},m_{b},\mu_{F})\, f_{a/A}(\xi,\mu_{F})\label{CxF}\end{eqnarray}
 into a convolutions of the Wilson coefficient functions ${\mathcal{C}}_{j/a}(x,\mathrm{b},m_{b},\mu_{F})$
and $k_{T}$-integrated parton distributions $f_{a/A}(\xi,\mu_{F})$.
The Sudakov exponential and ${\rm b}$-dependent parton densities
resum contributions from soft and collinear multi-parton radiation,
respectively. $Y\equiv\mbox{PERT}-\mbox{ASY}$ is the difference between
the finite-order cross section (PERT) and its asymptotic expansion
in the small-$q_{T}$ limit (ASY).

The Higgs cross sections depend on the mass $m_{b}$ of the bottom
quark. The distributions $\overline{{\mathcal{P}}}_{j/A}(x,{\rm b},m_{b})$
for the heavy quarks ($j=c,b$) cannot be reliably evaluated at all
impact parameters if a conventional factorization scheme, such as
the zero-mass variable-flavor number (ZM-VFN, or massless) scheme,
is used. The reason is that $\mHQ$ acts as an additional large momentum
scale, which, depending on the value of ${\rm b}$, introduces large
logarithms $\ln^{n}(\mHQ{\rm b})$ or non-negligible terms $\propto(\mHQ{\rm b})^{n}$.
The situation encountered here is reminiscent of the heavy-quark contributions
to the DIS structure functions $F_{i}(x,Q^{2}),$ which are not adequately
described by the conventional factorization schemes at either small
or large momentum transfers $Q^{2}$ (see, for instance, \cite{Tung:2001mv}).
To work around this complication, Ref.~\cite{Nadolsky:2002jr} proposed
to formulate the CSS formalism in a general-mass variable flavor number
(GM-VFN) scheme~\cite{Collins:1998rz}, which correctly evaluates
the heavy-quark mass effects at all momentum scales. Among all GM-VFN
factorization schemes, the S-ACOT scheme \cite{Collins:1998rz,Kramer:2000hn}
was found to be well-suited for the efficient calculation of the CSS
resummed cross sections. In particular, in this heavy-quark CSS (CSS-HQ)
formalism \cite{Nadolsky:2002jr} the dependence on $m_{b}$ is dropped
in all ${\cal O}(\alpha_{s})$ terms in Eq.~(\ref{WYDY}) except
for $\overline{{\mathcal{P}}}_{b/A}(x,{\rm b},m_{b})$. 

\begin{figure}[tb]
\begin{center}\includegraphics[%
  clip,
  width=0.49\columnwidth]{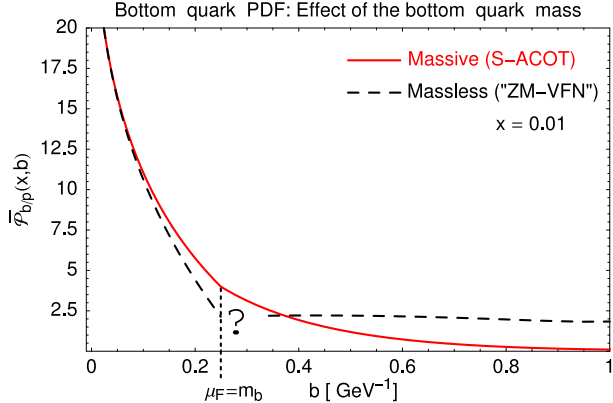}\end{center}

\caption{The bottom-quark distributions $\overline{{\mathcal{P}}}_{\qHQ/p}(x,{\rm b},\mHQ)$
in the proton vs. the impact parameter~${\rm b}$. The solid and
dashed curves correspond to the S-ACOT and massless ({}``ZM-VFN'')
factorization schemes, respectively.\label{fig:PbA}}
\end{figure}

The dependence of the bottom-quark parton density $\overline{{\mathcal{P}}}_{\qHQ/p}(x,{\rm b},\mHQ)$
on the impact parameter is shown in Fig.~\ref{fig:PbA}. The ZM-VFN
parton density $\overline{{\mathcal{P}}}_{\qHQ/p}(x,{\rm b},\mHQ)$
is not properly defined below the threshold $\mu_{F}=\mHQ$ (or above
${\rm b}=b_{0}/\mHQ$). It was continued to large ${\rm b}$ in the
previous calculations using an effective {}``ZM-VFN'' approximation
described in Ref.~\cite{Berge:2005rv}. The S-ACOT parton density
$\overline{{\mathcal{P}}}_{\qHQ/p}(x,{\rm b},\mHQ)$ is well-defined
at all ${\rm b}.$ It reduces to the ZM-VFN result at ${\rm b}\ll b_{0}/\mHQ$
and is strongly suppressed at ${\rm b}\gg b_{0}/\mHQ$. The suppression
is caused by the decoupling of the heavy quarks in the parton densities
at $\mu_{F}$ much smaller than $\mHQ$ (${\rm b}$ \nolinebreak
much larger than $b_{0}/\mHQ$). Consequently the impact of the non-perturbative
contributions from ${\rm b}\gtrsim1\mbox{\, GeV}^{-1}$ is reduced
in the heavy-quark channels compared to the light-quark channels. 

\begin{figure}
\includegraphics[%
  scale=0.4]{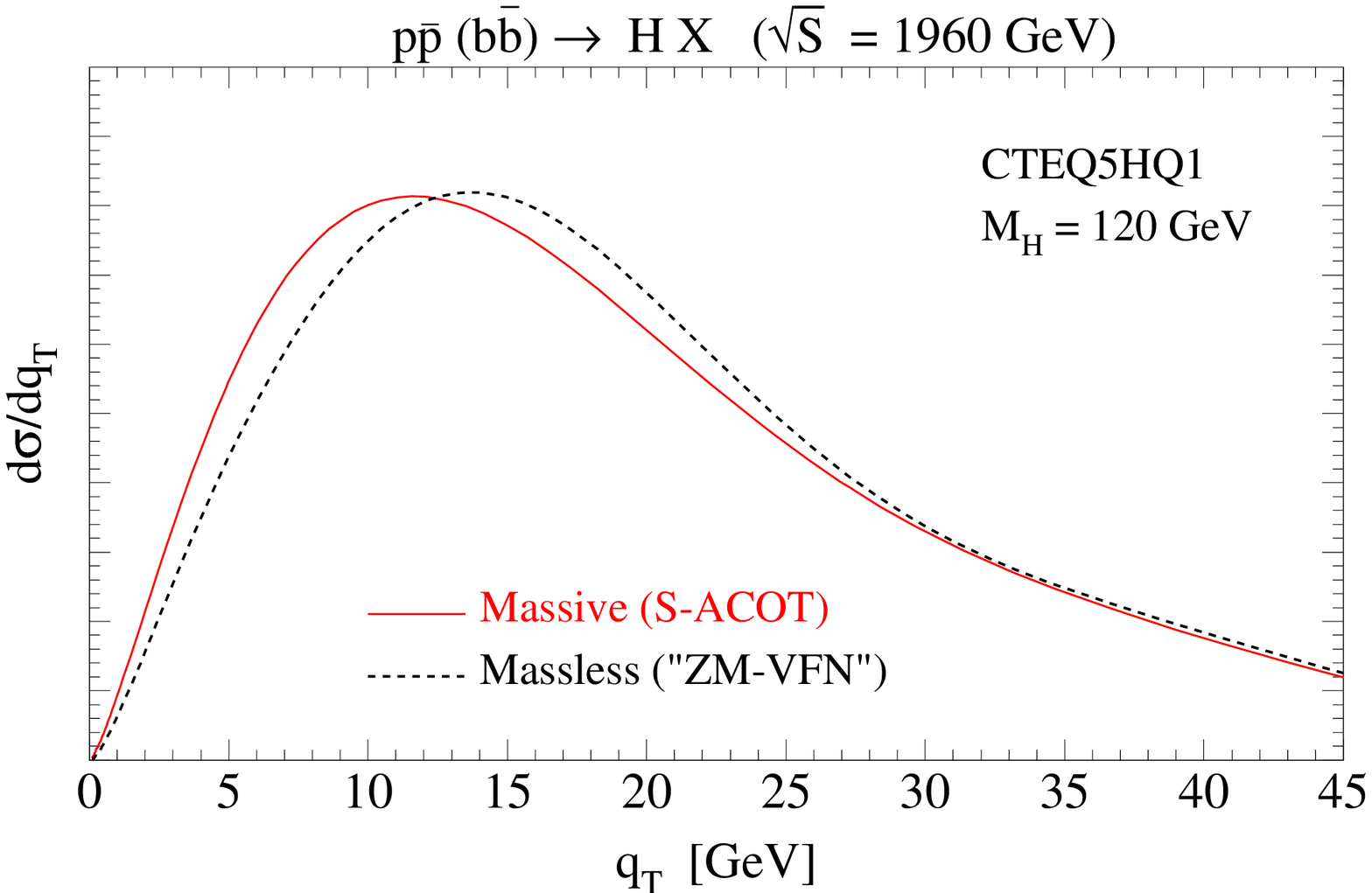}\includegraphics[%
  scale=0.4]{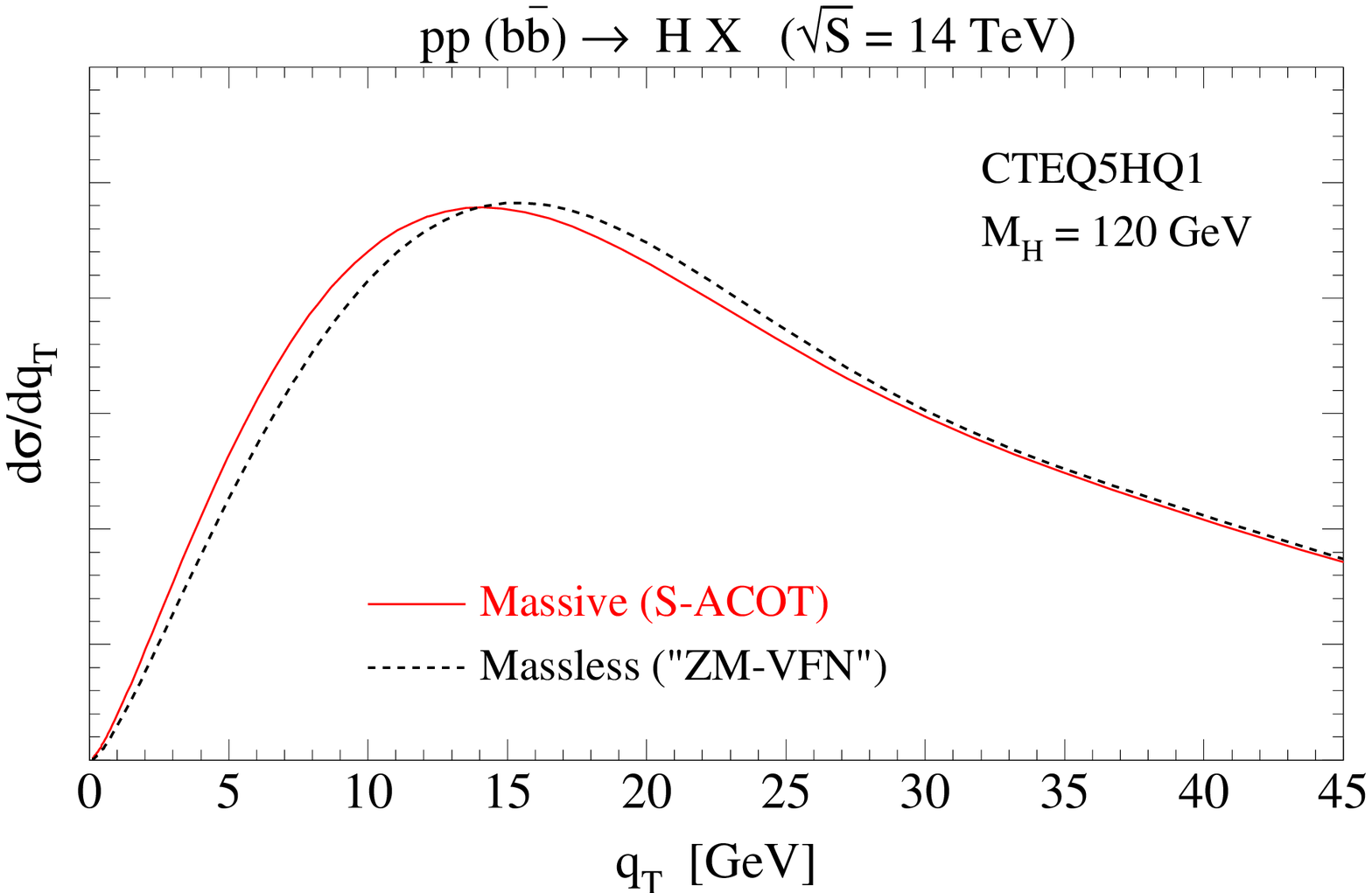}\vspace*{-17pt}

\begin{center}\hspace*{1.cm}(a)\hspace*{7.5cm}(b)\vspace*{-22pt}\end{center}

\caption{Transverse momentum distribution of on-shell Higgs bosons in the
$b\bar{b}\to\mH$ channel at (a)~the Tevatron and (b)~LHC. The solid
(red) lines show the $q_{T}$ distribution in the massive (S-ACOT)
scheme. The dashed (black) lines show the distribution in the massless
({}``ZM-VFN'') scheme. The numerical calculation was performed using
the programs Legacy and ResBos~\cite{Balazs:1997xd,Landry:2002ix}
with the CTEQ5HQ1 parton distribution functions~\cite{Lai:1999wy}.
The bottom quark mass is taken to be $m_{b}=4.5$~GeV.\label{fig:bbh_lhc}}
\end{figure}

The massless ({}``ZM-VFN'') calculation therefore underestimates
the true behavior at ${\rm b}>\nolinebreak0.1\mbox{ GeV}^{-1}$ and
small $q_{T}$. This effect can be seen in Fig.~\ref{fig:bbh_lhc},
which displays $d\sigma/dq_{T}$ for $b\bar{b}\rightarrow\mH$~boson
production at (a)~the Tevatron and (b)~LHC.%
\footnote{Fig.~\ref{fig:bbh_lhc} does not specify the overall normalization
of $q_{T}$ distributions. It is valid for both Standard Model and
supersymmetric Higgs bosons, since at leading order the supersymmetric
result can be obtained by rescaling the Standard Model $b\bar{b}H_{SM}$
coupling: $g_{b\bar{b}\{ h,H,A\}}^{MSSM}=\{-\sin\alpha,\cos\alpha,\sin\beta\,\gamma_{5}\} g_{bbH}^{SM}/\!\cos\beta$.
The net effect of $m_{b}$ on $q_{T}$ distributions will be the same
for the SM and MSSM Higgs bosons, up to an overall normalization constant. %
} At the Tevatron, the $q_{T}$ maximum shifts in the {}``ZM-VFN''
approximation to larger $q_{T}$ by about $2$~GeV out of $11.7$~GeV
(about $17\,$\%). For a Higgs mass $M_{H}=200$~GeV, the maximum
of $d\sigma/dq_{T}$ shifts by about 1.9 GeV out of $12.7$~GeV.
At the LHC, the difference between the {}``ZM-VFN'' and S-ACOT calculations
is smaller compared to the Tevatron, because the influence of the
${\rm b}>0.1\,\mbox{GeV}^{-1}$ region is reduced at smaller momentum
fractions~$x$ probed at the LHC~\cite{Qiu:2000hf}. The maximum
of the $q_{T}$ distribution shifts in the {}``ZM-VFN'' approximation
by about $1.3$~GeV (9\% out of $14.1$~GeV) to larger $q_{T}$.
The results for other Higgs masses can be found in Ref.~\cite{Berge:2005rv}.

\section{Numerical Comparison with PYTHIA\label{sec:Numerical-results}}

\begin{figure}
\begin{center}\includegraphics[%
  width=0.50\textwidth]{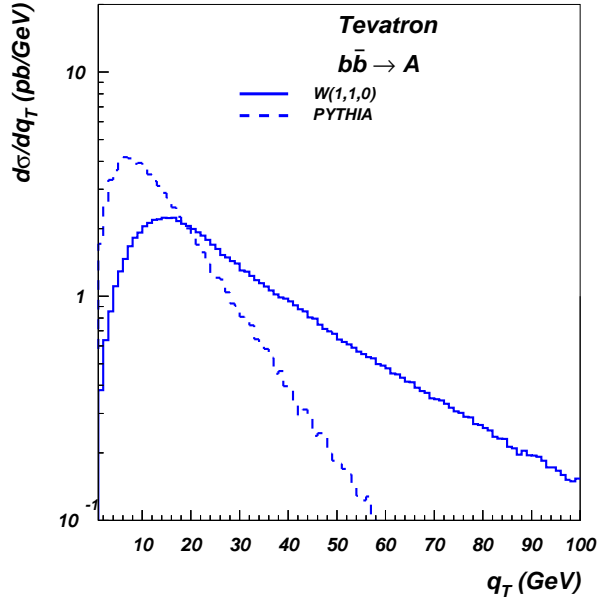}\end{center}

\caption{$q_{T}$ distributions for production of 100 GeV CP-odd Higgs bosons
$A$ via $b\bar{b}$ fusion in the Tevatron Run-2. The solid and dashed
curves correspond to the lowest-order $W$-term $W(1,1,0)$ (with
functions ${\cal A}(\alpha_{s}(\bar{\mu}))$ and ${\cal B}(\alpha_{s}(\bar{\mu}))$
evaluated at ${\cal O}(\alpha_{s})$) and PYTHIA. \label{Fig:one}}
\end{figure}

The full $q_{T}$ dependence of the $b\bar{b}\rightarrow\mH$ process
is affected by constraints on phase space available for QCD radiation
(less relevant at small $q_{T}$). We illustrate the interplay of
various effects by comparing the CSS-HQ resummation to the PYTHIA
Monte Carlo program~\cite{Sjostrand:2001yu}. We focus on production
of the CP-odd Higgs particle $A$ for $\tan\beta=50$ (predictions
for the other Higgs bosons can be obtained by rescaling the $b\bar{b}A$
coupling).

As compared to the CSS-HQ formalism, the PYTHIA calculation does not
include contributions generated from the ${\cal C}$-functions and
$Y$-term, and it evaluates the soft parton contributions at ${\cal O}(\alpha_{s}).$
Therefore, we start by comparing the PYTHIA $q_{T}$ distribution
to the resummed $W$-term $W(1,1,0)$ in Eq.~(\ref{WYDY}), with
the functions ${\cal A},$ ${\cal B}$, and ${\cal C}$ in Eqs.~(\ref{Sudakov}),
(\ref{CxF}) being evaluated at orders $\alpha_{s},$ $\alpha_{s}$,
and $\alpha_{s}^{0}$, respectively. The orders of $\alpha_{s}$ in
${\cal A},$ ${\cal B}$, and ${\cal C}$ are shown as the arguments
of $W(1,1,0)$. 

It is evident from Fig.~\ref{Fig:one} that the shapes of $W(1,1,0)$
and PYTHIA $q_{T}$ distribution are very different, though the integrated
rates (\emph{i.e.,} the areas under the two curves) are about the
same. The $q_{T}$ distribution from PYTHIA is narrower and peaks
at lower $q_{T}$ than $W(1,1,0)$. The large discrepancy between
the two curves is in contrast to the case of $W$ and $Z$ production
via light-quark scattering, where the above two calculations predict
similar, though not identical, $q_{T}$ distributions~\cite{Balazs:1997xd}. 

A closer examination reveals that additional features must be implemented
in the resummed cross section in order to reliably describe the $q_{T}$
distributions of Higgs bosons produced via $b\bar{b}$ fusion. 

\begin{itemize}
\item The kinematical effects account for a large part of the disparity
between $W(1,1,0)$ and PYTHIA. The bottom-quark PDF is a rapidly
decreasing function of $x$ in the probed range of $x$. Consequently,
approximations for the true partonic kinematics (especially those
made for the light-cone momentum fractions $x$) may have a strong
impact on the rate of $b\bar{b}$ scattering. This feature should
be contrasted to the behavior of the light-quark PDF's in $W$ and
$Z$ production, which include a substantial valence component and
vary slower with $x$. As a result, the kinematical approximations
are less consequential in the $W$ and $Z$ case.\\
{\small ~}\\
When PYTHIA generates QCD radiation, the kinematical distributions
of the final-state particles, including the quarks and gluons from
the QCD showering, are modified to satisfy energy-momentum conservation
at each stage of the showering. In the resummation calculation, information
about the exact parton kinematics is included in the finite-order
term (PERT). The resummed cross section is therefore expected to be
closer to PYTHIA once the ${\cal O}(\alpha_{s})$ finite term, PERT(1)-ASY(1),
is included. In the $W(1,1,0)$ calculation, the emitted gluons are
assumed not to carry any momentum at all in the soft limit. To compensate
for small, but nonzero energy of the soft gluon emissions, we introduce
a {}``kinematical correction'' (KC) in the W and ASY terms. This
correction modifies the minimal values of partonic momentum fractions
$x_{A}$ and $x_{B}$ in order to account for reduction of phase space
available for collinear QCD radiation at large $q_{T}$. 
\item The lowest-order cross section $W(1,1,0)$ does not evaluate effects
of the bottom-quark mass, which is first included in the ${\cal C}$-function
of order $\alpha_{s}$. Also, additional, though not complete, $O(\alpha_{s}^{2})$
contributions arise in the Sudakov form factors inside PYTHIA when
the next-to-leading order PDF's are used. To account for both features,
we evaluate the $W$ term at one higher order (2,2,1) and include
the $m_{b}$ dependence using the CSS-HQ scheme. 
\end{itemize}
\begin{figure}
\begin{center}\includegraphics[%
  width=0.50\textwidth]{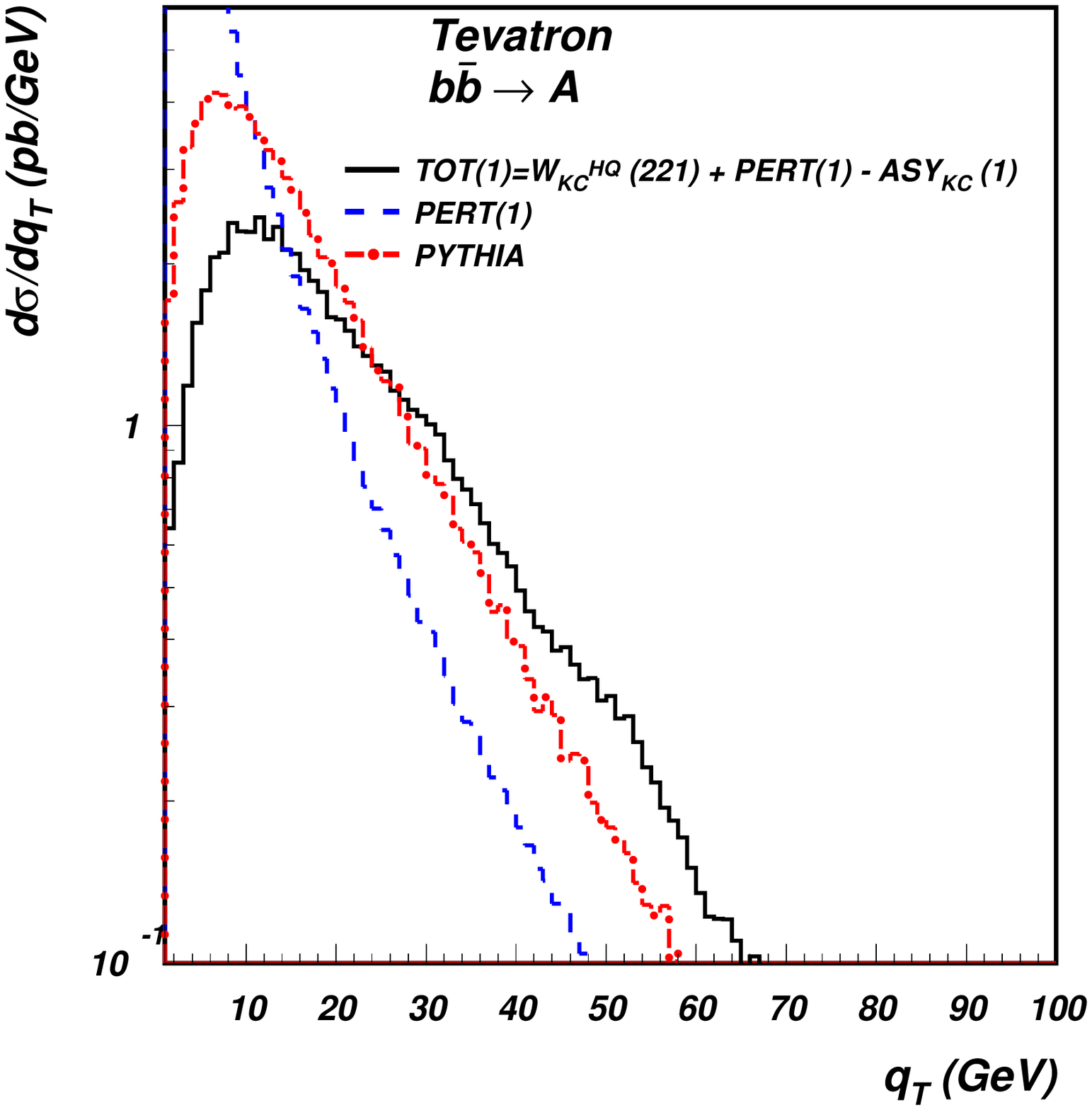}\includegraphics[%
  width=0.50\textwidth]{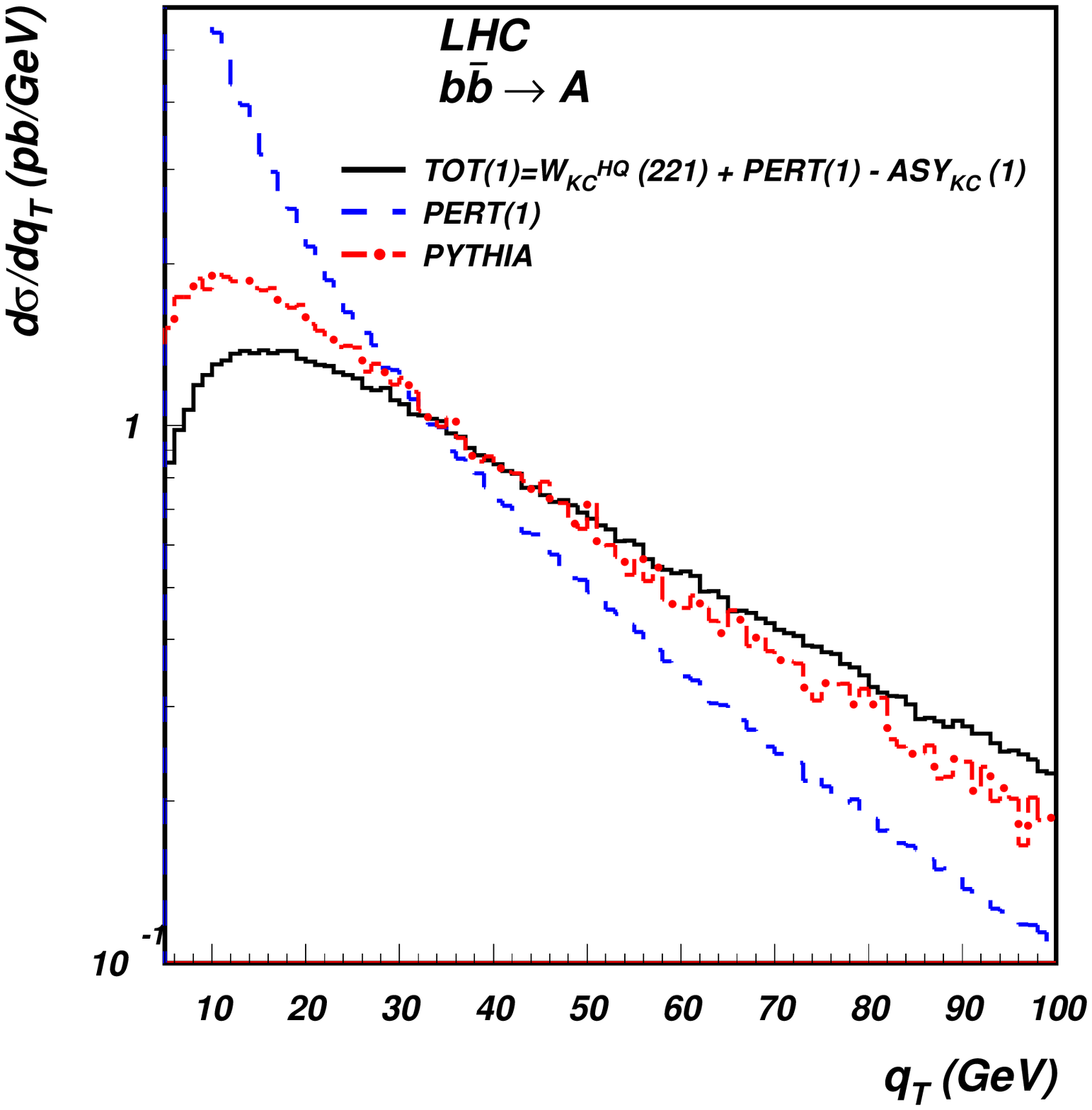}\end{center}

\begin{center}(a)\hspace{3in}(b)\end{center}

\caption{Comparison of $q_{T}$ distributions predicted by TOT(1), PERT(1)
and PYTHIA, for Higgs boson produced via $b\bar{b}$ fusion at (a)
the Tevatron Run-2 and (b) LHC, for $M_{A}=$100 and 300 GeV respectively.
\label{fig:three}}
\end{figure}

Thus, our full prediction TOT(1) is obtained by adding ${\textrm{W}}_{\textrm{KC}}^{\textrm{CSS-HQ}}$(2,2,1)
(evaluated in the CSS-HQ formalism with the kinematical correction)
and PERT(1), and subtracting ${\textrm{ASY}}_{\textrm{KC}}$(1). It
is shown for $M_{A}=100$~GeV at the Tevatron in Fig.~\ref{fig:three}(a)
and $M_{A}=300$~GeV at the LHC in Fig.~\ref{fig:three}(b). TOT(1)
(solid line) is compared to the fixed-order prediction PERT(1) (dashed)
and the PYTHIA prediction (dot-dashed). As one can see, the results
for Tevatron and LHC are qualitatively similar. TOT(1) is closer to
the PYTHIA prediction than $W(1,1,0)$, though the two distributions
are not identical. The PYTHIA $q_{T}$ distribution peaks at lower
$q_{T}$ than TOT(1). In the large $q_{T}$ region, the TOT(1) rate
is larger than the PYTHIA rate.

\begin{figure}
\includegraphics[%
  width=0.50\textwidth]{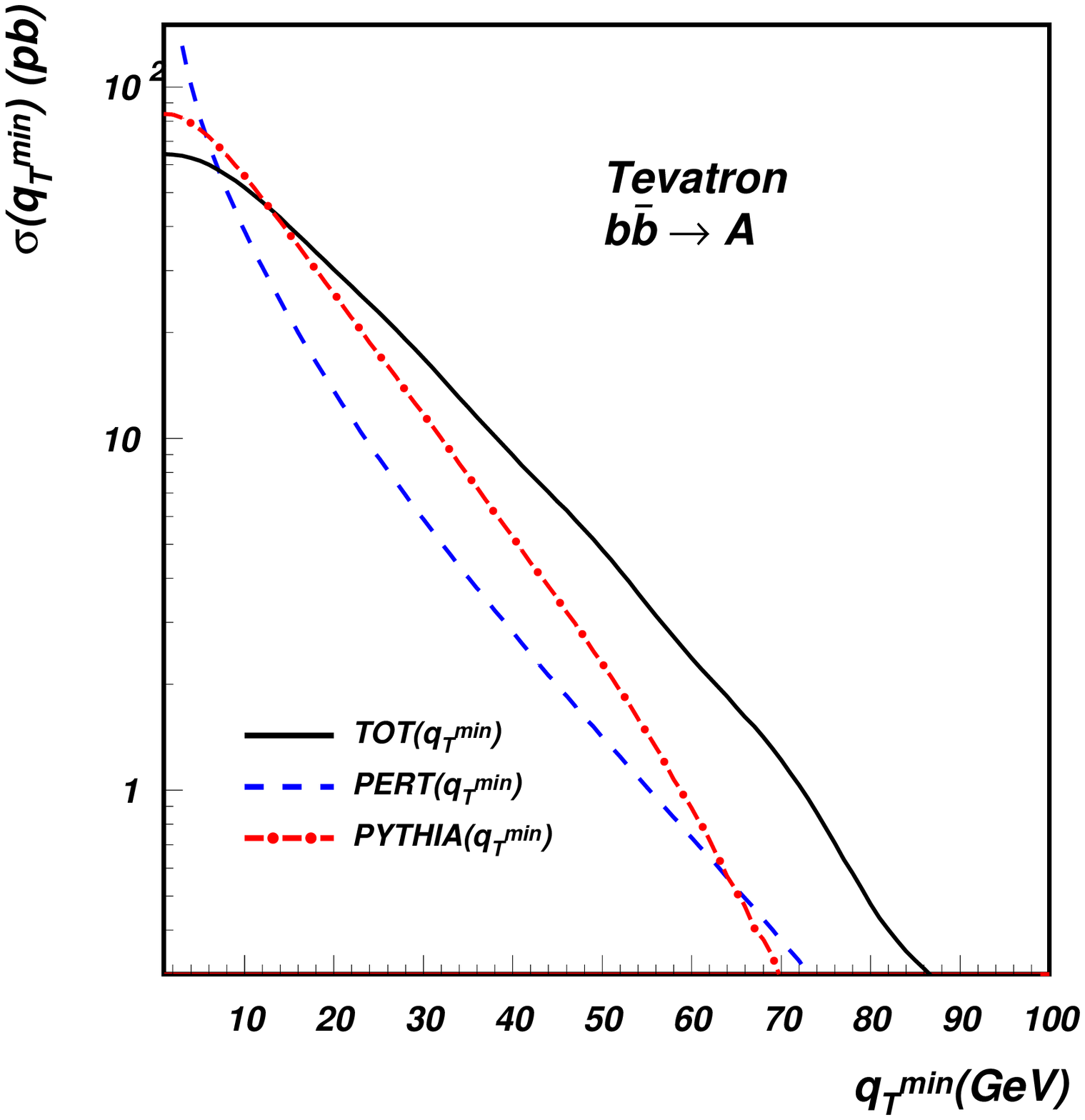}\includegraphics[%
  width=0.50\textwidth]{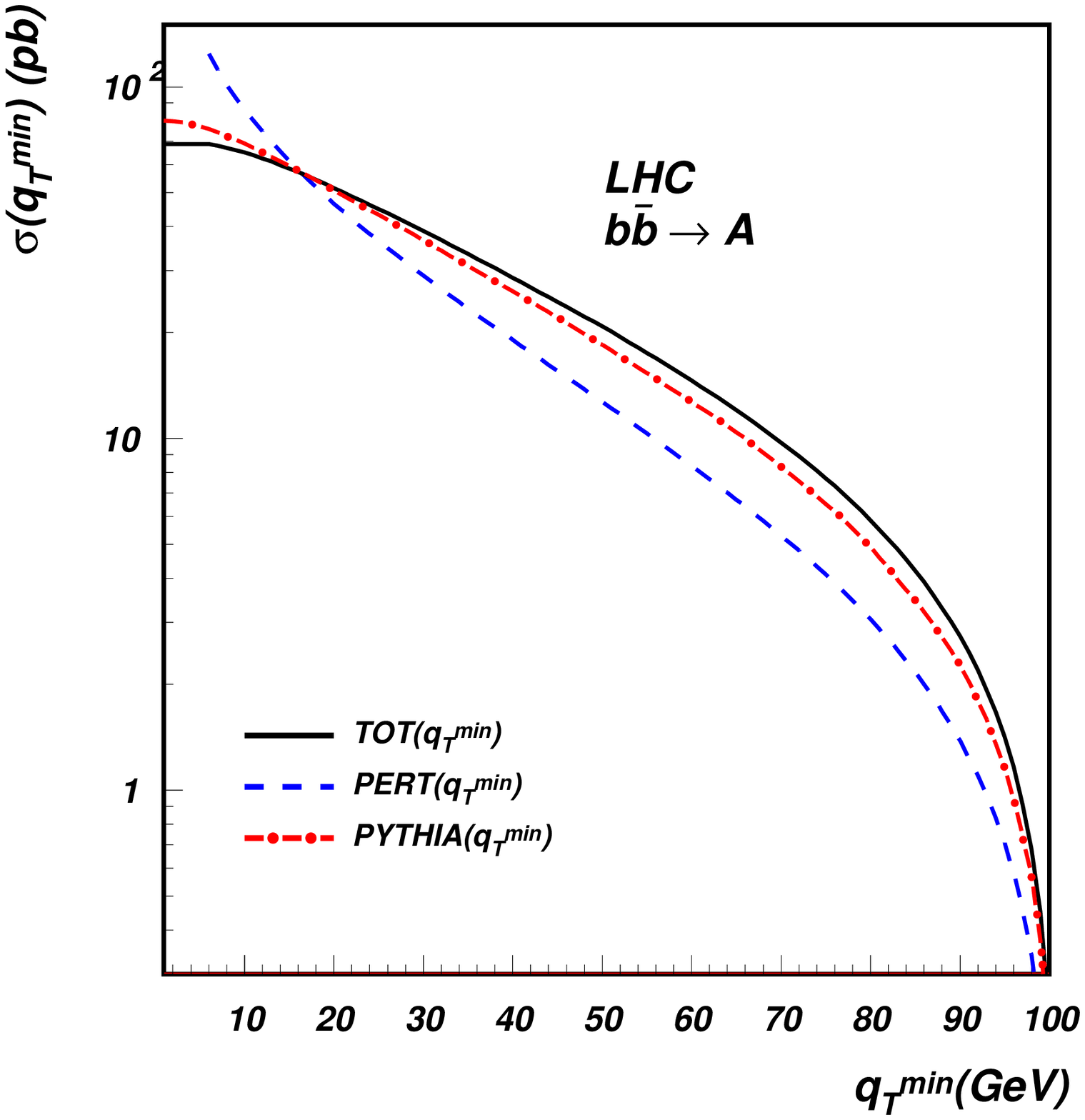}

\caption{Comparison of the integrated rates, deduced form Fig.~\ref{fig:three},
as a function of the minimal $q_{T}$ value taken in the integration
over $q_{T}$ at the Tevatron Run-2 (left) and LHC (right) for $M_{A}=$100
and 300 GeV, respectively. \label{fig:four} }
\end{figure}

Finally, Fig.~\ref{fig:four} shows the integrated cross section
as a function of the minimal $q_{T}$ in the calculation for the Tevatron
(left) and LHC (right). This is another way to illustrate the differences
in the shapes of $q_{T}$ distributions obtained in the resummation,
fixed-order, and PYTHIA calculations.

\section{Conclusion\label{sec:Conclusion}}

Multiple parton radiation in $b$-quark scattering is conspicuously
sensitive to effects of large bottom-quark mass $m_{b}$ and phase-space
constraints on collinear emissions. Both $m_{b}$ dependence and phase-space
dependence tangibly modify the shape of Higgs $q_{T}$ distributions
in the $b\bar{b}\rightarrow\mH$ processes. The two types of effects
were consistently implemented within the CSS resummation formalism
for heavy-quark scattering \cite{Nadolsky:2002jr,Berge:2005rv,Belyaev:2005bs},
realized in a massive (GM-VFN) factorization scheme. These corrections
act on different $q_{T}$ regions. When the dependence on $m_{b}$
is taken into account, the position of the peak in the $d\sigma/dq_{T}$
distribution shifts to a lower $q_{T}$ value, leaving the rate at
large $q_{T}$ essentially unchanged. The kinematical correction is
effective in the high-$q_{T}$ region, where it largely reduces the
Higgs production rate. As a result, we obtain an improved prediction
for the full $q_{T}$ spectrum of Higgs bosons, an important piece
of information needed for the future Higgs searches.

\section*{Acknowledgments}

This work was supported in part by the U.S. Department of Energy under
grant DE-FG03-95ER40908, contract W-31-109-ENG-38, and the Lightner-Sams
Foundation. We also acknowledge the support in part by the U.S. National
Science Foundation under awards PHY-0354838 and PHY-0244919.

%\bibliographystyle{apsrev}
%\bibliography{hbpt}

\hyphenation{Post-Script Sprin-ger}

\end{document}